\documentclass[%
 preprint,
superscriptaddress,
groupedaddress,
unsortedaddress,
altaffillsymbol,
 amsmath,amssymb,
 aps,
prc,
]{revtex4-1}

\usepackage{graphicx}
\usepackage{dcolumn}
\usepackage{bm}
\usepackage{url}

\begin{document}

\preprint{APS/123-QED}

\title{Low spin spectroscopy of neutron-rich $^{43,44,45}$Cl via $\beta^-$ and $\beta n$ decay}

\author{Vandana Tripathi$^{1}$\footnote{Corresponding author: vtripath@fsu.edu}} 
\author{Soumik Bhattacharya$^{1}$} 
\author{E.~Rubino$^{1,9}$} 
\author{C.~Benetti$^{1,3}$} 
\author{J.~F.~Perello$^{1}$} 
\author{S.~L.~Tabor$^{1}$}
\author{S.~N.~Liddick$^{2,3,4}$}
\author{P.~C.~Bender$^5$}
\author{M.~P.~Carpenter$^{6}$} 
\author{J.~J.~Carroll$^7$}
\author{A.~Chester$^{2,3}$}
\author{C.~J.~Chiara$^7$}
\author{K.~Childers$^{2,4}$}
\author{B.~R.~Clark$^{8}$}
\author{B.~P.~Crider$^8$}
\author{J.~T.~Harke$^9$}
\author{R.~Jain$^{3,10}$}
\author{B.~Longfellow$^{2,10,9}$}
\author{S.~Luitel$^8$}
\author{M.~Mogannam$^{3,4}$}
\author{T.~H.~Ogunbeku$^8$}
\author{A.~L.~Richard$^{2,9}$}
\author{S.~Saha$^5$}
\author{N.~Shimizu$^{11}$}
\author{O.~A.~Shehu$^8$}
\author{Y.~Utsuno$^{12,13}$}
\author{R.~Unz$^8$}
\author{Y.~Xiao$^8$}
\author{S.~Yoshida$^{14}$}
\author{Yiyi Zhu$^5$}

\affiliation{$^{1}$Department of Physics, Florida State University, Tallahassee, Florida 32306, USA}
\affiliation{$^2$National Superconducting Cyclotron Laboratory, Michigan State University, East Lansing, MI 48824, USA}
\affiliation{$^{3}$Facility for Rare Isotope Beams, Michigan State University, East Lansing, Michigan 48824, USA}
\affiliation{$^{4}$Department of Chemistry, Michigan State University, East Lansing, Michigan 48824, USA}
\affiliation{$^5$Department of Physics, University of Massachusetts Lowell, Lowell, Massachusetts 01854, USA}
\affiliation{$^{6}$Argonne National Laboratory, Argonne, Illinois 60439, USA}
\affiliation{$^7$U.S. Army Combat Capabilities Development Command Army Research Laboratory, Adelphi, Maryland 20783 USA}
\affiliation{$^{8}$Department of Physics and Astronomy, Mississippi State University, Mississippi State, MS 39762, USA}
\affiliation{$^{9}$Lawrence Livermore National Laboratory, Livermore, California 94550, USA}
\affiliation{$^{10}$Department of Physics and Astronomy, Michigan State University, East Lansing, Michigan 48824, USA}
\affiliation{$^{11}$Center for Computational Sciences, University of Tsukuba, Tennodai, Tsukuba 305-8577, Japan}
\affiliation{$^{12}$Advanced Science Research Center, Japan Atomic Energy Agency, Tokai, Ibaraki 319-1195, Japan}
\affiliation{$^{13}$Center for Nuclear Study, University of Tokyo, Hongo, Bunkyo-ku, Tokyo 113-0033, Japan}
\affiliation{$^{14}$Liberal and General Education Center, Institute for Promotion of Higher Academic Education, Utsunomiya University, Mine, Utsunomiya, Tochigi 321-8505, Japan}

\date{\today}

\begin{abstract}

$\beta^-$ decay of neutron-rich isotopes $^{43,45}$S, studied at the National 
Superconducting Cyclotron Laboratory is reported here. $\beta$-delayed $\gamma$ transitions 
were detected by an array of 
16 clover detectors surrounding the Beta Counting Station which consists
of a 40x40 Double Sided Silicon Strip Detector followed by a Single Sided Silicon 
Strip Detector. $\beta$-decay half-lives have 
been extracted for $^{43,45}$S by correlating implants and decays in the 
pixelated implant detector with further coincidence with $\gamma$ transitions in the 
daughter nucleus. The level structure of $^{43,45}$Cl is expanded by 
the addition of 20 new $\gamma$ transitions
in $^{43}$Cl and 8 in $^{45}$Cl with the observation of core excited negative-parity 
states for the first time. For $^{45}$S decay, a large fraction of the  $\beta$ decay 
strength goes to delayed neutron emission populating states in $^{44}$Cl 
which are also presented. Comparison of 
experimental observations is made to detailed shell-model calculations using 
the $SDPFSDG-MU$ interaction to highlight the role of the 
diminished $N=28$ neutron shell gap and the near degeneracy of the proton $s_{1/2}$ 
and $d_{3/2}$ orbitals on the structure of the neutron-rich Cl isotopes. 
The current work also provides further support to a  ground state spin-parity 
assignment of $3/2^+$ in $^{45}$Cl.

\end{abstract}

\maketitle


\newpage

\section{\label{sec:intro}Introduction}

$^{48}$Ca is a doubly magic nucleus, characterized by a spherical shape 
which is stabilized by the $Z=20$ and $N=28$ shell closures. However, 
down the isotonic chain, $^{44}$S displays clear evidence of shape mixing 
or co-existence by the presence of an isomeric low-lying $0^+_2$ state 
\cite{isomeric_0plus}. For $^{46}$Ar, an isotope between $^{48}$Ca and $^{44}$S  
the situation seems more nuanced. The systematics of $2^+_1$ energies in the 
Ar isotopic chain indicate a pronounced shell gap but there remains 
the question that shell model calculations, irrespective of the interaction 
used, predict a high $B(E2)$ value in contradiction with the high-lying $2^+$ state 
\cite{Coulex_46Ar, blongfellow_Siso} in $^{46}$Ar. Further, based on recent mass 
measurements of $^{46-48}$Ar the extracted one- and two-neutron shell
gaps indicate the presence of a persistent, but reduced empirical shell gap 
in $^{46}$Ar compared to the doubly magic $^{48}$Ca \cite{mass_46_48Ar}. 
The occurrence of spherical or deformed shapes in nuclei is intimately related to the 
nucleonic structure; for the exotic $N=28$ isotones, deformation sets in 
due to the melting of the $N=28$ shell gap along with the closing of the 
$Z=16$ spherical gap. Hence the understanding of the single particle 
structure in the vicinity of $^{46}$Ar is very important which appears to 
be the transition between doubly magic $^{48}$Ca and the 
more collective $^{44}$S. 

Spectroscopy of odd-odd and odd-even nuclei can provide valuable 
information to test the single particle structure and benchmark calculations. 
Neutron-rich Cl isotopes with proton holes in the $sd$ shell
approaching the $N=28$ magic number can shed light on the single particle 
structure in this interesting region of the chart of nuclides. However 
spectroscopic information for $A>40$ isotopes of Cl is rather limited 
due to the difficulty in producing those for investigation
\cite{Sorlin_43_45Cl,Gade_KClP_gs,Stroberg_43_46Cl}. The most exhaustive 
information comes from in-beam $\gamma$-ray spectroscopy following fragmentation 
performed at NSCL for $^{43-46}$Cl \cite{Stroberg_43_46Cl} where $\gamma$-$\gamma$ 
coincidences were used to generate experimental level schemes up to $\approx$ 2 MeV for
$^{43}$Cl, $^{45}$Cl, and $^{46}$Cl, although no firm spin assignments
could be made from that measurement. Very recently the first $\gamma$-ray 
spectroscopy of $^{47,49}$Cl was performed at the Radioactive Isotope Beam Factory 
(RIBF) with $^{50}$Ar projectiles impinging on a liquid hydrogen target to 
produce the exotic isotopes \cite{linh_47_49Cl}. Through the one-proton knockout reaction, 
a spin-parity, $J^\pi = 3/2^+$ was proposed for the ground state of $^{49}$Cl. 

For the odd-A Cl isotopes with $N> 20$ filling of neutrons in the  $f_{7/2}$ orbital 
may cause inversion between the $\pi s_{1/2}$ and $\pi d_{3/2}$ orbitals. 
The strong monopole proton-neutron interaction between the $\pi d_{3/2}$ 
and $\nu f_{7/2}$ orbits acts attractively on the $\pi d_{3/2}$ 
orbital, lowering its energy with respect to the $\pi s_{1/2}$  orbital 
when the $\nu f_{7/2}$ orbital is being filled for the $N\approx 28$ nuclei.
 It is anticipated that there will be a reversion back to the normal ordering  for 
larger neutron excess (neutrons filling the $\nu p_{3/2}$ orbit) 
where the neutron-proton interaction between the 
$\pi s_{1/2}$ and $\nu p_{3/2}$ orbits comes into play. These changes 
in the single particle ordering can impact properties of ground and 
excited states of nuclei. For nearby odd-A K isotopes ($Z=19$) the ground-state spin inversion of
$3/2^+$ to $1/2^+$  is observed for $A = 47-49$~\cite{Broda_49K,Gade_KClP_gs}. 
In the case of Cl isotopes, direct spin measurements have not been performed for the 
neutron rich isotopes, however a picture similar to K can be drawn from other experimental information. 

In $^{41}$Cl the ground state is favored to be $1/2^+$ ~\cite{41Cl,41Cl_prc66,41Cl_prc67}. 
This is based on an analogy with the lighter isotopes where the yrast $5/2^+_1$ level decays 
exclusively to the $3/2^+$ and not to the $1/2^+$ level. 
The $\beta$-decay of $^{41}$S to $^{41}$Cl~\cite{winger_43S} also supports a $1/2^+$ assignment, 
however $3/2^+$ cannot be ruled out completely based on these observations. 
For $^{43}$Cl strong evidence in favor of an inverted ground state of $1/2^+$ comes from the 
in-beam $\gamma$ spectroscopy following fragmentation where the angular distribution 
of the 330-keV transition which connects the ground state to the first excited state was measured \cite{Sorlin_43_45Cl}. 
The anisotropy of the 330-keV transition rules out that it is a decay from a $1/2^+$ state 
making the ground state spin  to be  $1/2^+$ instead.  However a
reverse conclusion is drawn from the $\beta$-decay study of $^{43}$Cl to $^{43}$Ar ~\cite{winger_43Cl}. 
That is based on 
the observed branching to the ground state of 28(10)\% with 5.81 
as a lower limit of the log{\it ft}. The 
ground state of $^{43}$Ar being $5/2^-$, this log{\it ft} could correspond to a First Forbidden (FF) $\beta$ 
transition but only if $^{43}$Cl ground state is $3/2^+$.  This is a much weaker argument in favor of the $3/2^+$ assignment  
as extracting ground state branching has its limitations.  The assignment of a $3/2^+$ to the ground state 
of $^{45}$Cl has more support than $^{41,43}$Cl. In Ref.~\cite{Stroberg_43_46Cl}, the life time of the 
lowest 130-keV transition was measured to be 470(60) ps yielding a $B(M1)_{exp} = 0.055\mu_N^2$ 
which is better reproduced by shell model calculation in that 
work for a $3/2^+$ ground state. Additionally in the one proton knockout from $^{45}$Cl 
to $^{44}$S by Riley et al,~\cite{lewriley_45Cl}, the population of the $4_1^+$ in $^{44}$S provides a strong argrument for 
a $3/2^+$ ground state in $^{45}$Cl. A $1/2^+$ ground state for $^{45}$Cl would lead to an 
unmeasurably small cross section for the $4_1^+$ state based on shell model calculation.  In our own 
recent study of the $\beta^-$-decay of $^{45}$Cl to $^{45}$S, a $3/2^+$ assignment was favored over 
$1/2^+$ based on the strong population of the $5/2^+$ state in the daughter~\cite{soumik_45Cl}. 
Moving to more exotic 
isotopes, a $J^\pi = 3/2^+$ was proposed for the ground state of $^{49}$Cl 
in the RIBF experiment~\cite{linh_47_49Cl} as mentioned before,
though the 
ground-state for $^{47}$Cl is not yet confirmed, but likely to be $3/2^+$.

In the odd-A Cl isotopes, the negative-parity states arise 
from the excitation of the unpaired proton or one of the $sd$ neutrons to
the $fp$ shell. In the stable $^{37}$Cl with $N=20$ a sequence of 
$7/2^-$-$9/2^-$-$11/2^-$-$13/2^-$ states is observed with the $7/2^-_1$ 
state at 3.1 MeV \cite{37Cl_negative,37Cl_negative_pol}. For $^{39}$Cl the 
negative-parity states occur at about half the excitation energy 
with a sequence of $5/2^-$-$7/2^-$-$9/2^-$-$11/2^-$ states
 starting at 1697 keV \cite{39_41Cl_negative}. 
 The information for more exotic $^{41}$Cl is sparse with only two 
 tentative negative-parity states proposed at 1475 keV 
 ($5/2^-,7/2^-$) and 2718 keV ($5/2^-,7/2^-$) \cite{39_41Cl_negative} 
 with none known for $^{43,45}$Cl.

The current work is an effort to enhance the spectroscopic information on exotic Cl isotopes and we have 
studied $\beta^-$ decay of $^{43,45}$S to access excited states in the 
daughters $^{43,44,45}$Cl. The $\beta$-decay half-life of $^{43}$S
has been reported earlier \cite{grevy,marek,winger_43S} with an evaluated value 
of 265(15) ms \cite{NDS_A43}, however none followed the $\gamma$ transitions in
the daughter nucleus. Similarly the literature value for $^{45}$S half-life is 68(2) ms 
\cite{grevy,sorlin_Pn} from implant-$\beta^-$ correlations performed at GANIL. 
The current work, to our knowledge, will be the first to report on 
$\beta$-delayed $\gamma$ transitions in the daughter nuclei $^{43,44,45}$Cl to 
complement the limited information from in-beam $\gamma$ spectroscopy. 
For the $N=26$ $^{43}$Cl 20 new transitions have been identified while 8 new ones 
have been added for the $N=28$ $^{45}$Cl. $\beta^-$ decay from the $N=27,29$ isotopes 
of $S$ which have negative-parity ground states will feed 
negative-parity intruder states via Gamow-Teller (GT) $\beta$ transitions, reported 
here for the first time. 

The experimental results are interpreted within configuration interaction Shell Model (SM) 
calculations using the $SDPFSDG-MU$ interaction within a model space of the sd-pf-sdg 
shells \cite{utsuno, utsuno_new} with some truncation on $1p1h$ excitations across 
either the $Z=N=20$ or $N=40$ shell gaps. Protons are allowed to 
move only from the $sd$ to the $pf$ shell while one neutron excitations from 
$sd$ to $pf$ and $pf$ to $sdg$ (for $N> 20$) shells are explicitly taken into account.
The valence space contains the $sd$ orbitals for protons and the $fp$ orbitals for
neutrons, with a $^{16}$O inert core. The interactions used for the sd shell, pf shell,
and the cross shell are respectively, USD, GXPF1B, and a variant of the 
$V_{MU}$ that was employed for the $SDPF-MU$ interaction \cite{utsuno}. 
The calculations 
predict very close-by $3/2^+$ and $1/2^+$ states with one being the ground state for 
odd-A Cl isotopes. Closer to stability $3/2^+$ is the preferred ground state which switches to 
$1/2^+$ for $^{41}$Cl due to the near degeneracy of $s_{1/2}$ and $d_{3/2}$ proton 
orbitals together with enhanced collectivity.

\section{\label{sec:exp}Experimental Setup}

The experiment  was carried out at the National  Superconducting  Cyclotron Laboratory
(NSCL) \cite{brad} at Michigan State University where a $^{48}$Ca primary beam was 
fragmented to produce several exotic isotopes of P, S and Cl with neutron number 
around 28. Some interesting results from this data can be found in 
Refs. \cite{tripathi_new, soumik_45Cl} and further details of the setup can also 
be found there. The primary beam was accelerated 
to 140 MeV/u and then fragmented  on a thick $^9$Be target at  the  target  position  
of  the fragment separator, the A1900 \cite{A1900}, and further dispersed through 
a wedge-shaped Al degrader positioned at the intermediate dispersive image of the A1900. 
$^{43}$S and $^{45}$S were produced using two different settings of the 
A1900 with $1\%$ and $2\%$ momentum acceptance respectively. 

In the two cases $^{43}$S and $^{45}$S isotopes (along with others) 
were implanted in a pixelated 
(40 strips x 40 strips) Double-Sided Silicon Strip Detector (DSSD), which is 
part of the Beta Counting Setup (BCS). An Al degrader placed upstream ensured that 
the implants stopped roughly at the middle of the 1mm thick DSSD. 
The DSSD was followed by a Single-Sided Silicon Strip Detector (SSSD) which 
served as a veto detector for light ions passing through. A low implant rate of 
about 150/s ensures clean correlations between the implants and decay events. 
Two Si PIN detectors placed upstream of the DSSD provided the energy loss and 
time of flight information  which, along with  the scintillator  at the intermediate 
dispersive image of the A1900 were used for particle  identification of  the 
incoming implants at the  BCS. The DSSD and SSSD stack was surrounded by 16 
Clover detectors to  record the $\beta$-delayed $\gamma$ rays with a total singles 
efficiency of about $5\%$ at 1 MeV after addback. Energy and efficiency calibration 
was performed using standard $\gamma$ sources up to 3.5 MeV. The data were collected 
event by event using the NSCL digital data acquisition system  \cite{prokop}. 
Each channel provided its own time-stamp signal, which allowed  coincidences and 
correlations to be built in the analysis offline.

\begin{figure}
	\includegraphics[width=\columnwidth]{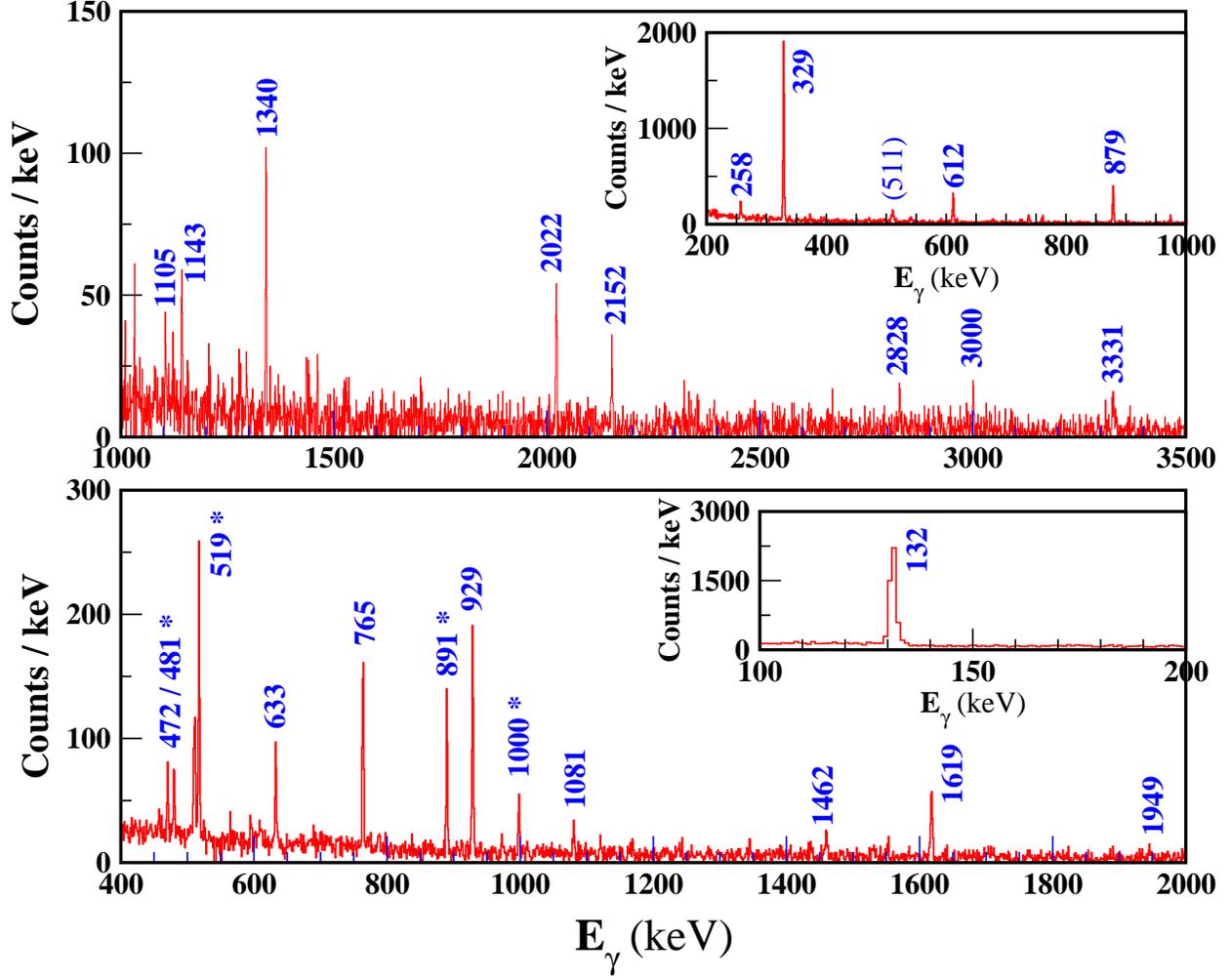}
	\caption{\label{fig:43S_45S_gammas}
		Top: $\beta$-delayed $\gamma$ transitions in $^{43}$Cl for a correlation time
		window of 200 ms ($\approx 1~ T_{1/2}$) after subtracting a background spectrum generated from a 
		1000 ms correlation window. 
		Bottom: Similarly,  $\beta$-delayed $\gamma$ transitions in $^{45}$Cl for a correlation 
		window of 100 ms ($\approx 1~ T_{1/2}$) after subtracting a background spectrum generated from a 
		1000 ms correlation window. 
		For both the strong transitions are labeled while the weaker ones are displayed in the coincidence 
		spectra (Figure~\ref{fig:43S_coincidence} and  Figure~\ref{fig:45S_coincidence}). Transitions 
		marked with an asterisk in the bottom panel are in the $\beta 1n$ daughter, $^{44}$Cl.
	}
\end{figure}

 $\gamma$ transitions coincident with $\beta$ correlated implants for $^{43}$S and $^{45}$S
decays are shown in the top and bottom panel of Figure~\ref{fig:43S_45S_gammas} 
respectively. For both a correlation window 
roughly equal to the half-life of parent was chosen to showcase the 
$\beta$-delayed $\gamma$ transitions in the daughter. Random correlations were avoided 
by subtracting $\gamma$ transitions associated with a very long correlation window. 
Most of the strong transitions in the daughter nucleus are clearly observed 
and later the $\gamma$-$\gamma$ coincidences 
will be discussed which allowed us to build the level schemes.

\section{\label{sec:results}Experimental Results}

\subsection{\label{43S} $^{43}$S (Z=16; N=27) $\beta^-$ decay}

In a typical continuous beam $\beta$-decay experiment with fast fragmentation
beams the half-life extraction requires event by event time correlation between 
$\beta$ particles and the precursor implants. Such time and spatial 
correlation was followed in the current analysis to generate decay curves. 
Figure~\ref{fig:43S_decaycurve} (a) shows the decay curve for the $^{43}$S 
implants for a correlation time window of 2 seconds. 
The figure also shows the fit following Bateman equations yielding a value of 
256(5) ms for the half-life of $^{43}$S similar to the  
evaluated value of 265(15) ms from prior studies \cite{NDS_A43}. 
In the fit the contribution of $\beta 1n$ daughter {\it i.e.} the $P_n$ value
was taken to be 40(10) \% \cite{sorlin_Pn, marek}. 
For further confirmation, the half-life was also obtained  by fitting the time 
distribution  of the $\gamma$-ray transitions at 329- and 879 keV in the 
$\beta 0n$ daughter as shown in Figure~\ref{fig:43S_decaycurve} (b). 
An exponential fit with a constant  background yields a half-life of 
250.2(25) ms. According to SM calculations discussed earlier using the 
$SDPFSDG-MU$ interaction \cite{yoshida_website}, predictions for the $^{43}$S 
half-life with a $3/2^-$ ground state are 336 ms when including First Forbidden 
(FF) $\beta$ transitions and 
360 ms when considering only GT transitions. Both these values are 
longer than the measured value, possibly correlated to the prediction of a 
low $P_n$ value of only 8\% by the shell model calculations \cite{yoshida_website}. 

\begin{figure}
	\includegraphics[width=\columnwidth]{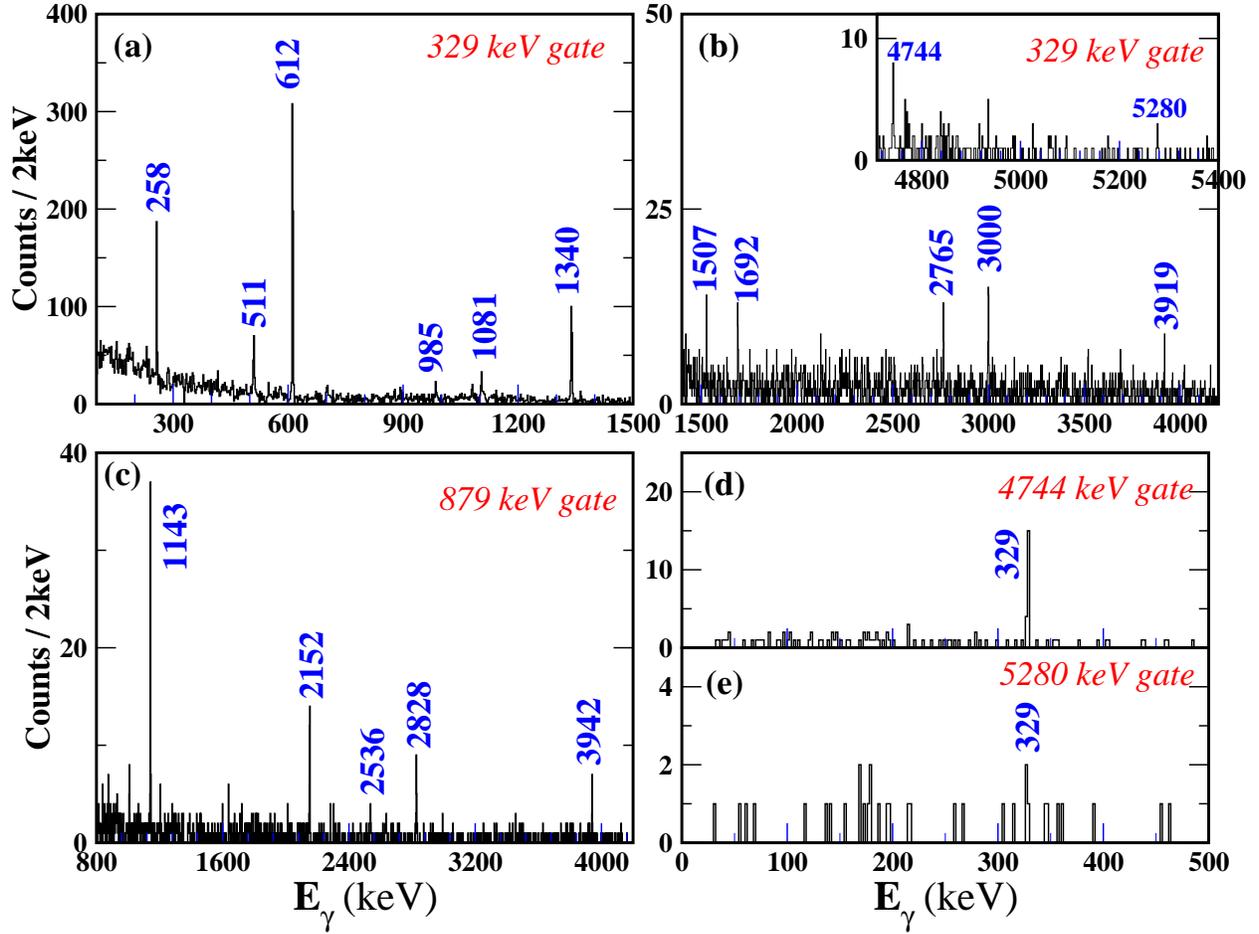}
	\caption{\label{fig:43S_coincidence}
		(a)-(b) $\gamma$ transitions in $^{43}$Cl observed in coincidence with the 
		 329-keV $\gamma$ transition which de-excites the $3/2^+_1$ state to the 
		ground state. The inset of (b) shows the high energy $\gamma$ transitions (c) spectrum showing the $\gamma$ transitions coincident with 
		the 879-keV transition. (d)-(e) the 329-keV transition is seen in coincidence 
		with the high lying 4744- and 5280-keV transitions.
	}
\end{figure}

The level-scheme of $^{43}$Cl derived from the current delayed $\gamma$-ray 
spectroscopy is shown in Figure~\ref{fig:43S_levelscheme}. 
The energy levels up to 1927 keV with 6 $\gamma$ transitions 
were known from the prior studies \cite{Stroberg_43_46Cl,Sorlin_43_45Cl,41Cl}. 
We have been able to extend the level-scheme to 5612 keV with 20 newly identified 
$\gamma$-ray transitions, shown in red. Placement of all the newly observed $\gamma$-ray 
transitions were confirmed with $\gamma$-$\gamma$ coincidences. 
Figure \ref{fig:43S_coincidence} shows the $\gamma$-ray peaks in coincidence with 
the previously-known ground-state transitions at 329 keV (a,b) and 879 keV (c). 
The 879 keV level with a direct decay to the ground state was only tentatively placed 
in Ref. \cite{Stroberg_43_46Cl}. Prior to that, the 881 keV (879 in this work) was 
proposed to depopulate a 1830-keV level feeding into the 941 keV state 
\cite{Sorlin_43_45Cl}. However being the second strongest transition and showing 
no coincidence with the 329-keV transition (Figure~\ref{fig:43S_coincidence}), we 
propose that the 879-keV transition comes from an excited state at the same energy 
ratifying the tentative placement in Ref. \cite{Stroberg_43_46Cl}. 
The decay of the 2022 keV level to both the 879-keV and 329-keV states, 
among others, confirms further the placement of the 879-keV transition. The placement 
of the 1509-keV transition  from Ref. \cite{Sorlin_43_45Cl} (1507 keV in this work)
is also 
confirmed by its coincidence with the 329-keV transition. The coincidence of the 
highest energy $\gamma$ peaks 4744 and 5280 keV with the 329 keV transition 
are shown in panels (d) and (e) in the same figure. Table \ref{tab:43Cl_gamma} 
gives the details of all the $\gamma$ transitions where the absolute intensity 
was calculated using the measured peak area, $\gamma$ detection efficiency 
and the number of 
implants obtained from the fit to the decay curve. The log$ft$ values for the excited 
states above the 1927 keV level which we believe are directly populated in the $\beta$ decay 
via $GT$ transitions were calculated using the NNDC log$ft$ calculator 
\cite{nndc_logft} making use of the measured $T_{1/2}$ and absolute $\beta$ branching 
and the known $Q_{\beta^-}$ value~\cite{Q_beta}.

\subsection{\label{45S} $^{45}$S (Z=16; N=29) $\beta^-$ decay}

The decay curves for $^{45}$S decay with and without coincidence with $\gamma$ 
transitions in the daughter nucleus are shown in Figure~\ref{fig:45S_decaycurve}. 
The decay curve from the implant-$\beta$ correlations was fitted similar to 
$^{43}$S and a half-life of 69(1) ms was extracted in excellent agreement with 
the previous measurement from Ref. \cite{grevy}. Figure~\ref{fig:45S_decaycurve} (b) 
shows the time distribution of the 132 keV lowest transition in $^{45}$Cl and an 
exponential fit yields a half-life of 74(5) ms, the higher error bar is due to lower 
statistics. Contrary to the case of $^{43}$Cl the calculated half-life from the 
SM calculation after including both the GT and FF transitions at 
70.7 ms agrees well with the measured value. Excluding the FF transitions 
predicts 78.5 ms, slightly longer than the measured value. This is under the 
assumption that the ground state of $^{45}$S is $3/2^-$ which is expected from the 
filling of the $\nu p_{3/2}$ orbital for the $N=29$ isotope.

\begin{figure}
	\includegraphics[width=\columnwidth]{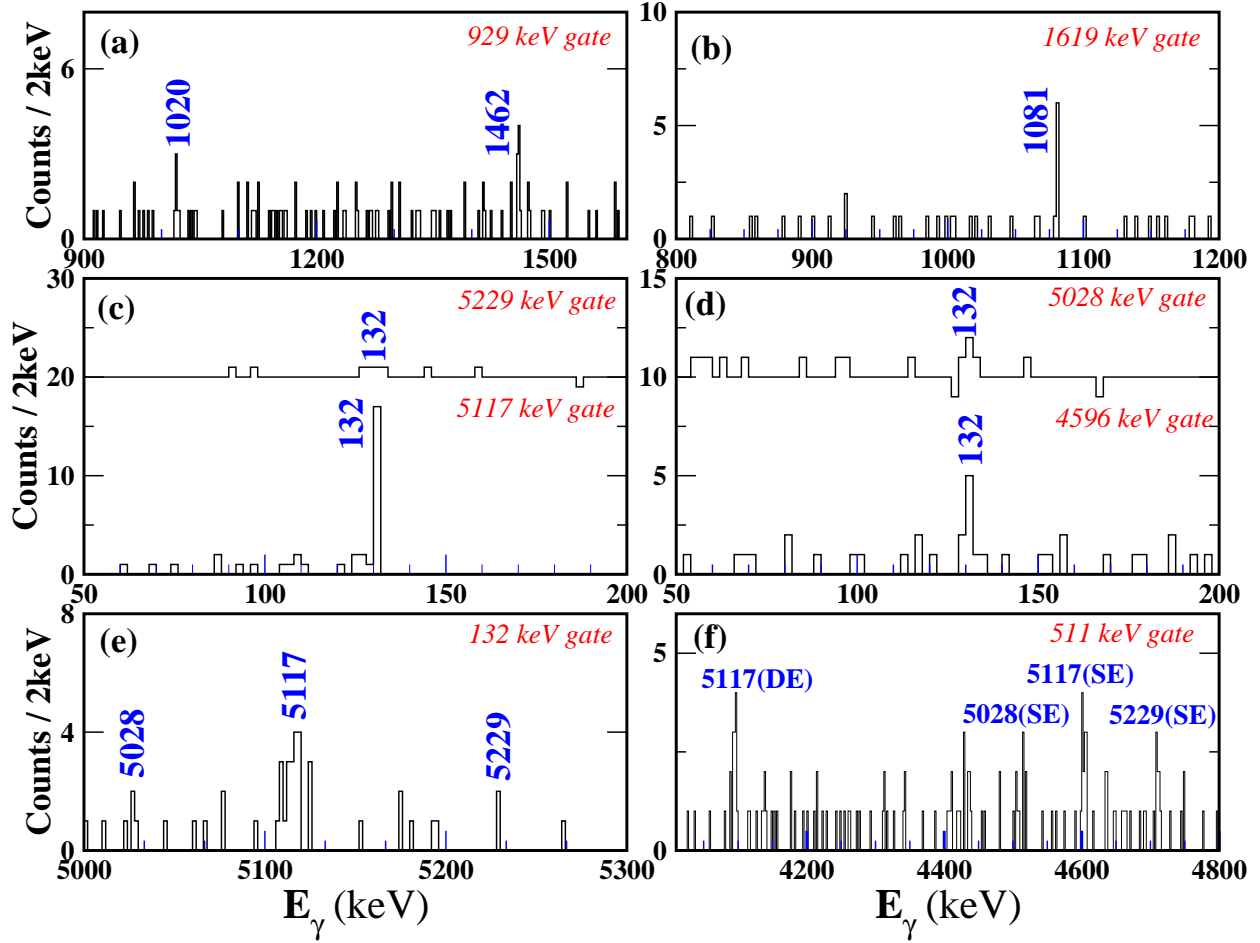}
	\caption{\label{fig:45S_coincidence}
		Spectra displaying the coincidences observed between the $\gamma$ transitions 
		in $^{45}$Cl.
		(a) the 1020- and 1462- keV transitions are shown in the 929-keV gate.
		(b) coincidences observed between the 1619-keV and 1081-keV transitions.
		(c)-(d) The 132-keV transition is seen in the gates of the transitions at 
		4596-, 5028-, 5118-, and 5229 keV. The y-scales for the 5028- and 
		5229 keV gates were shifted upwards by 10 and 20 counts respectively for 
		clarity of display. (e) The 5028-, 5118-, and 5229 keV transitions are 
  seen in the 132-keV gate. (f) the gate on the 511 keV $\gamma$ transition from pair production 
  in the detector which displays the single (SE) and double escape (DE) peaks of the 
  high energy $\gamma$ transitions as noted.
	}
\end{figure}

The $\beta$-delayed $\gamma$ transitions were studied to construct the 
level scheme of daughter $^{45}$Cl. The known $\gamma$ transitions 
at 132-, 633-, 765-, 929-, and 1619 keV \cite{Stroberg_43_46Cl} were 
clearly observed and are shown in black in Figure~\ref{fig:45S_levelscheme}. 
The 132-, 633-, 765-, and the 929 keV transitions were also observed in the 
$\beta$-delayed one-neutron emission from $^{46}$S published earlier by us 
\cite{tripathi_new}. 
Accurate energy values for these transitions are determined in this work as 
detailed in  Table \ref{tab:45Cl_gamma}. 
Whenever possible $\gamma$-$\gamma$ coincidences were explored to place the 
observed transitions, and some of the coincidences  are shown in 
Figure~\ref{fig:45S_coincidence}. A 1081 keV transition is seen in clear 
coincidence with the 1619 keV decay and hence proposed to feed that level. 
In Ref. \cite{Stroberg_43_46Cl} where states in $^{45}$Cl were investigated 
following fragmentation, a 1061(9) keV transition was placed above the 
1619 keV transition, however we do not observe that transition. Instead our 
observation is of the 1081 keV $\gamma$ ray. As the new transition is outside of 
error bar on the 1061 keV transition we have shown it as red in 
Figure~\ref{fig:45S_levelscheme} though the two could likely be the same transition 
with the $\beta$ decay spectrum giving a determination of energy closer to the 
true value. 

Coincidences are also observed between the 1462 keV and 929 keV 
transitions and hence 1462 is placed feeding the 929 keV level in departure from 
Ref. \cite{Stroberg_43_46Cl} where a transition of the same energy was considered 
a ground state transition. In the same work a 1950 keV transition was observed 
(tentative) but not placed in the level scheme. Based on coincidences observed 
between the 929 keV and 1020 keV transitions as seen in 
Figure~\ref{fig:45S_coincidence} (a) we determined a level at 1949 keV. 
The direct decay from  the level is also present but is partly obscured 
by a long lived activity line at 1945 keV and hence its intensity 
cannot be determined.  Beyond these previously known transitions, 
4 $\gamma$ transitions with energies higher than 4 MeV were observed. All of them, 
namely 4596-, 5028-, 5117-, and 5229 keV transitions are in 
coincidence with the 132-keV ground state transition. The reciprocal 
coincidences are shown in Figure~\ref{fig:45S_coincidence} (c,d) while 
Figure~\ref{fig:45S_coincidence} (e) shows the 132-keV gate where the $\gamma$ 
transitions above 5 MeV are shown. 
The strongest of this set is the 5117-keV for which we could also observe the 
first and second escape peaks. A gate on the 511 keV annihilation photon in 
Figure~\ref{fig:45S_coincidence} (f) shows the escape peaks for the highest energy 
$\gamma$ transitions. 

The absolute intensities of all the 
$\gamma$ transitions in $^{45}$Cl are given in Table~\ref{tab:45Cl_gamma}. 
The level scheme from the current work shown in Figure~\ref{fig:45S_levelscheme} 
displays the absolute $\beta$ feeding branches calculated using the intensities of the 
$\gamma$ rays and the total implants obtained from the fit to the decay curve.
The branching to the low-lying positive parity states especially to the 
$1/2^+_1$ and $3/2^+_2$ seems to indicate a significant branch due to FF decay. 
The SM calculations predict log$ft$ values of 6.83, 6.95 and 6.84 for the FF 
transitions to the $1/2^+_1$ and $3/2^+_{1,2}$ states respectively~\cite{yoshida_website} which would be 
consistent with observed feeding of the $1/2^+_1$ and $3/2^+_2$ states though in the 
current level scheme the feeding to the ground state is zero. 
Another source of large feeding to the low lying states has been speculated to be
feeding from $\gamma$-decay of unbound states as noted in Ref~\cite{gottardo}.
The states at 4728-, 5160-, 5249-, and 5361 keV are assumed to be directly 
fed by GT transitions and the log$ft$ values are calculated accordingly. 

\subsection{\label{45S} $^{45}$S (Z=16; N=29) $\beta 1n$ decay}

The large $Q_{\beta^-}$ value of 14.92(33) MeV for $^{45}$S \cite{Q_beta} 
leads to a
high $\beta$-delayed neutron emission probability ($P_n$). We 
followed the $\gamma$ activity of grand-daughter nuclei in $\beta 0n$, 
$\beta 1n$ and $\beta 2n$ descendants. The efficiency corrected 
intensities of the most intense $\gamma$ transitions in $^{45,44,43}$Ar 
were used to determine 
the $P_n$ value to be 60(10) \% for $^{45}$S. The error bar takes into 
account the correction of growth and decay of activities and the errors 
in branching ratios. This delayed neutron emission is consistent 
with $\beta$ feeding to the bound states of 45(5) \% which would be 
the lower limit as the level could be incomplete due to unobserved 
$\gamma$ transitions. The previous measurement of $P_n$ value from 
Ref.~\cite{sorlin_Pn} is 54\% (errors not quoted in the paper) in agreement 
with our value. The SM calculations predict a value of 40\%, lower than 
our measured value.

Due to the large $\beta$-delayed neutron strength, we were able to observe $\gamma$ 
transitions in $^{44}$Cl, the $\beta 1n$ daughter. The level scheme proposed 
by Stroberg {\it et al}, \cite{Stroberg_43_46Cl} from in-beam $\gamma$ spectroscopy 
following fragmentation could be verified with the addition of the 
481 keV $\gamma$ decay from the 999-keV state feeding the 518 keV state supported 
by $\gamma$-$\gamma$ coincidences. Though Ref. \cite{Stroberg_43_46Cl} 
did not make any spin-parity assignments to the excited states in $^{44}$Cl 
we proposed them to have negative parity as they are 
populated via the $\beta 1n$ channel of $^{45}$S $\beta$ decay which has a ground state 
$J^\pi$ = $3/2^-$ as tentatively marked in Figure~\ref{fig:44Cl_levelscheme}. 
The 891- and 999-keV transitions were also observed in the $\beta$-decay 
of $^{44}$S \cite{tripathi_new} where they would be populated via FF transitions. 
Specific spin parity assignments will be discussed in the next section.

\section{\label{discus}Discussion}

\subsection{\label{positive} $^{43,45}$Cl: Positive parity states}

The ground state of $^{43}$Cl is largely assumed to be $1/2^+$ as discussed in the 
introduction and the first excited state is 
considered a $3/2^+$ corresponding to the odd proton occupying either the $1s_{1/2}$ or 
the $0d_{3/2}$ orbital as per the shell model calculations. The shell model calculations 
performed for this work further predict a sequence of 
$1/2^+(gs)$-$3/2^+$-$3/2^+$-$5/2^+$-$5/2^+$ states for $^{43}$Cl 
(Figure~\ref{fig:43S_levelscheme}). 
The newly established 879-keV level is thus a candidate for the $3/2^+_2$ state 
while the 941-keV level is proposed as $5/2^+_1$ consistent with Ref.~\cite{Sorlin_43_45Cl}. 
We assign the 1669-keV level as the second $5/2^+$ 
state following the predicted sequence where no assignment was made to this level
earlier~\cite{Stroberg_43_46Cl}. 
The next level at 1836 keV is suggested to have $J^\pi = 7/2^+$. 
A level at a close by energy of 1830(8) keV was observed in Ref.~\cite{Sorlin_43_45Cl}
and proposed as a $7/2^+$ state. The $\gamma$ transition from that level was 881 keV 
which could be the 894 keV transition decaying to the $5/2^+_1$ as seen in this work.
We also observed the 1507 keV transition from the 1836 keV level to the $3/2^+_1$ 
state consistent with a $7/2^+$ assignment. The 1927 keV state is a candidate for the 
$7/2^+_2$ state predicted at a similar energy with a strong decay to the $5/2^+_2$
state.

Shell model predictions for $^{45}$Cl are also a sequence of 
$1/2^+(gs)$-$3/2^+$-$3/2^+$-$5/2^+$-$5/2^+$ states as the lowest excited states. 
However in our recent publication \cite{soumik_45Cl} following the 
$\beta^-$-decay of $^{45}$Cl to $^{45}$Ar, we made the argument for a $3/2^+$ 
ground state of the parent based on the strong population of $5/2^+$ state in $^{45}$Ar. 
Support for a $3/2^+$ ground state for $^{45}$Cl was also provided from the earlier
 observation of strong population of the $4^+_1$ state in $^{44}$Cl 
 from $^{45}$Cl knockout in Ref.~\cite{lewriley_45Cl}. 
Thus the first five experimental states in $^{45}$Cl are proposed as  
$3/2^+(gs)$-$1/2^+$-$3/2^+$-$5/2^+$-$5/2^+$ (Figure~\ref{fig:45S_levelscheme}) 
mostly similar to Ref.~\cite{Stroberg_43_46Cl}. However the state 
assigned as $5/2^+_2$ in Ref.~\cite{Stroberg_43_46Cl} is incorrect now due to the 
revised placement of the 1462 keV transition as noted before. 
The levels at 2391- and 2700 keV are 
proposed as the $7/2^+_1$ and $7/2^+_2$ states striking a similarity with the 
observation in $^{43}$Cl as also revealed in the calculations. The predicted 
 $7/2^+$ states at 2349 and 2629 would be good counterparts (Figure~\ref{fig:45S_levelscheme}).
  The 1949 keV level can be a candidate for the predicted 2132 keV, $1/2^+$ state which 
 would be consistent with its decay to the $5/2^+_1$ and $3/2^+_1$ states. 
 In Ref.~\cite{Stroberg_43_46Cl} also they had proposed a $1/2^+$ state above the 
 $5/2^+_2$ state.

\subsection{\label{positive} $^{43,45}$Cl: Ground state doublet}

\begin{figure}
	\includegraphics[width=\columnwidth]{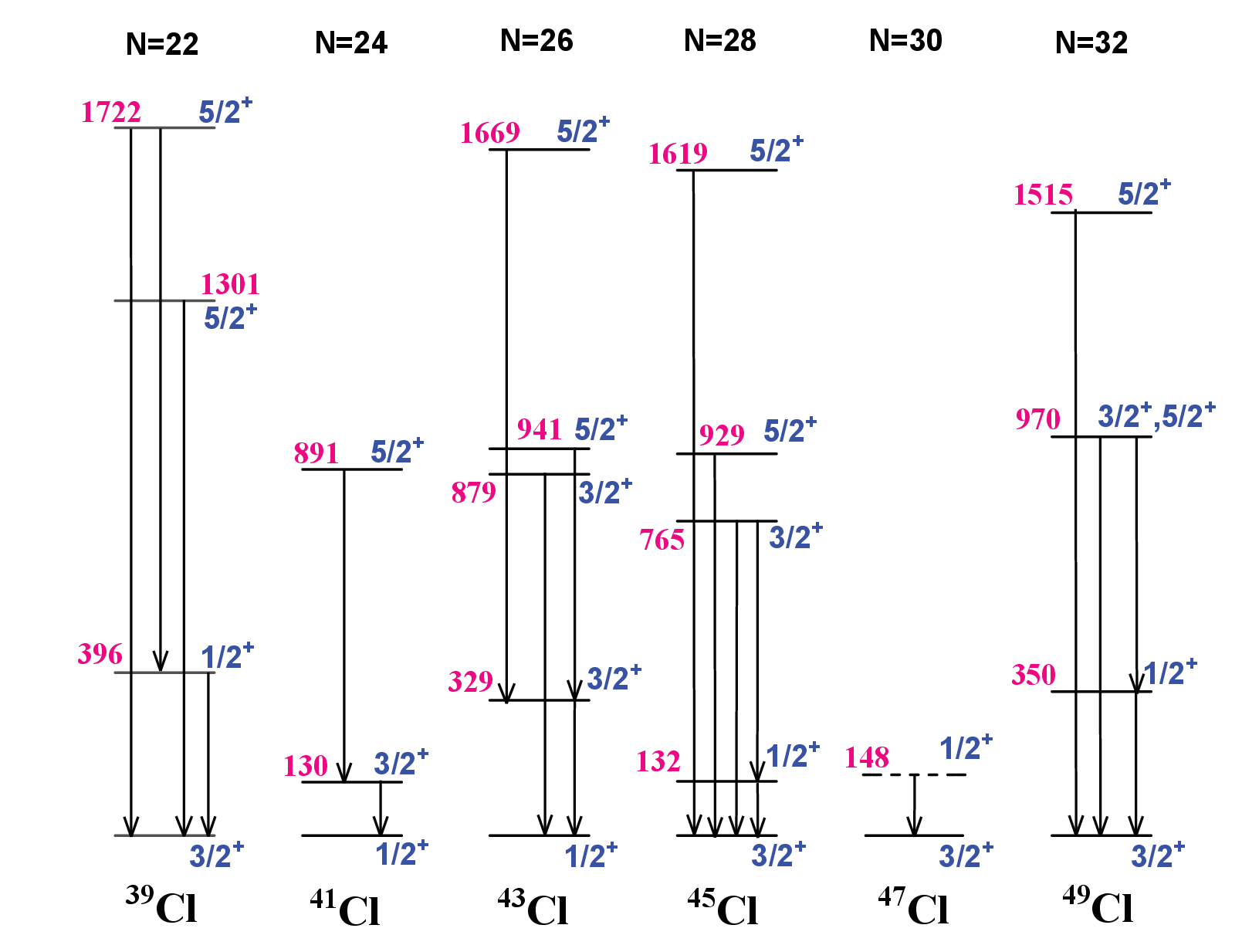}
	\caption{\label{fig:oddA_Cl_isotopes}
		Comparison of low-lying level structure of odd-A Cl isotopes from 
		$N=22$ to $N=32$. The data for $^{43,45}$Cl is from the current data set  
		while those of the other isotopes is from published work \cite{nndc}. 
		Though some of the spin parity assignments are tentative we did not put 
		them in parenthesis for ease of comparison. The point to emphasize here is the 
		strong decay branch for the $5/2^+$ states to the $3/2^+_1$ state and 
		not to the $1/2^+$ state. The only deviation is for $^{39}$Cl where both the 
		branches from the $5/2^+_2$ state have comparable strength. 
		This is also supported by SM calculations as shown in Table~\ref{tab:Cl_iso_branching}.
	}
\end{figure}

The ground state spin/parity of odd-A Cl isotopes is determined by the occupancy 
of the odd 
proton in either the $1s_{1/2}$ or $0d_{3/2}$ orbital and the lowest transitions of 
positive parity correspond to the excitation of the proton within the $sd$ shell. 
In Figure~\ref{fig:oddA_Cl_isotopes} we display the experimentally observed lowest 
excited states and $\gamma$ transitions in Cl isotopes from $N=22$ to $N=32$ as the 
neutrons fill up the lower part of the $fp$ shell. 
Closer to the shell gap at $N=20$ the ground 
state is $3/2^+$ with a well separated $1/2^+$ excited state corresponding to a hole 
in the $1s_{1/2}$ orbital. Away from $N=20$ the spacing between the $3/2^+$ and 
$1/2^+$ states is reduced reflective of the near degeneracy of the two orbitals as
neutrons fill up the $f_{7/2}$ orbital. Based on experimental observation till now 
a $1/2^+$ is favored for $^{41,43}$Cl whereas a $3/2^+$ assignment is better 
aligned with various experimental observations. 
The level structure of $^{45}$Cl in the current study further strengthens the 
proposed $3/2^+$ assignment based  on the stronger decay of the $5/2^+_1$ and 
$5/2^+_2$ states to the $3/2^+_1$ state relative to the $1/2^+_1$ state as was 
also seen in $^{39,41,43}$Cl.  
Predictions of the SM calculations for the decay of these $5/2^+$ states is given in 
Table~\ref{tab:Cl_iso_branching} where the trend of strong decays to the 
$3/2^+$ state can be seen. Consistent results are obtained using both experimental 
and theoretical energies for the excited states. 
Thus in $^{45}$Cl, with the observation of 
strong decay of the $5/2^+$ states exclusively to the ground state via the 930- and 
1619 keV transitions (no coincidence seen with the 132 keV transition), the assignment 
of $3/2^+$ to the ground state and $1/2^+$ to the 132 keV state seems more likely.

\begin{figure}
	\includegraphics[width=\columnwidth]{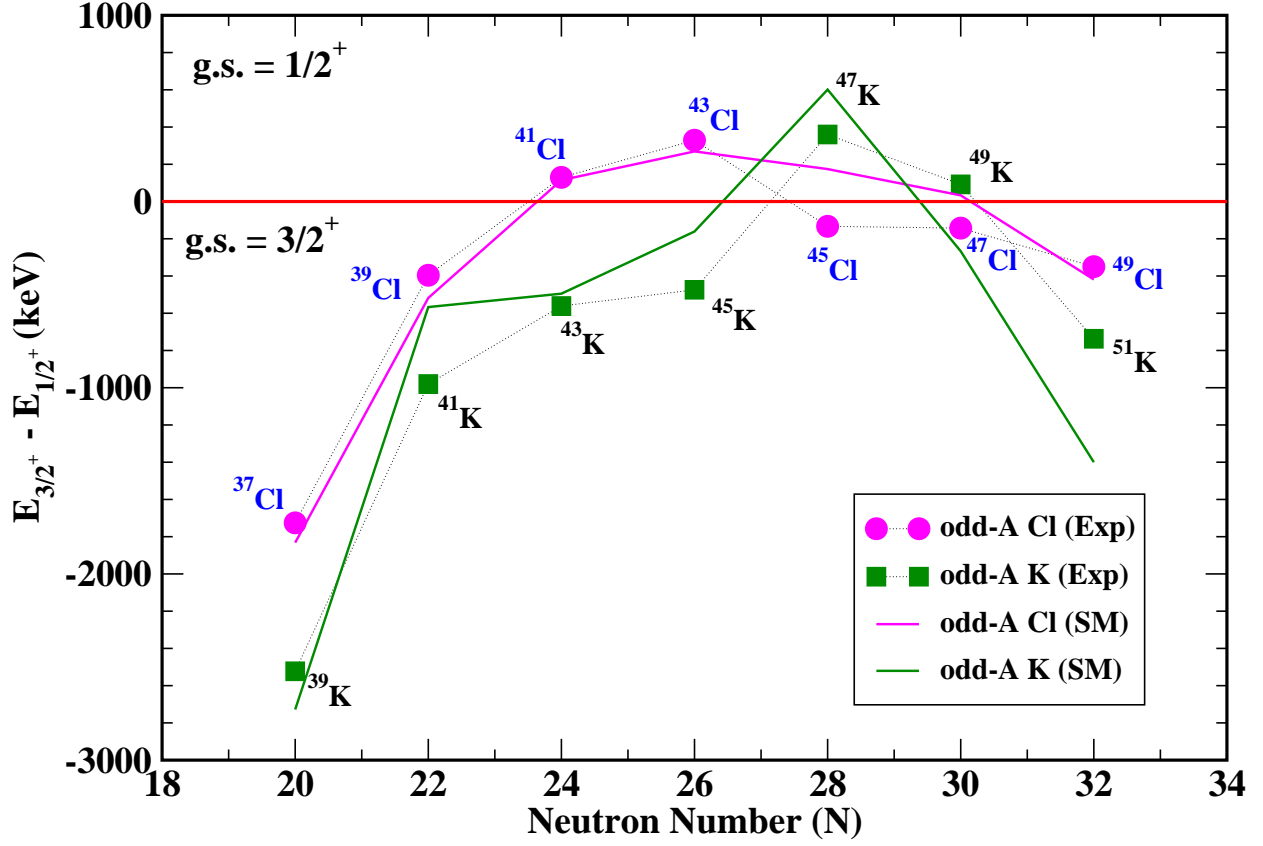}
	\caption{\label{fig:oddA_Cl_K_isotopes}
		The energy difference between the first $3/2^+$ and $1/2^+$ states 
		(one of them is the ground state) in odd-A Cl and K isotopes with neutron 
		numbers from 20 to 32 as observed in experiments is shown by the 
		solid symbols connected by thin dashed lines (magenta for Cl and green for K). 
		An inversion of the ground state $J^\pi$ is seen for both Cl and K for certain neutron numbers. 
		The data for $^{43,45}$Cl is from the current data set 
		while those of the other isotopes is from published work \cite{nndc}. 
		The solid lines are the corresponding energy 
		differences between the $3/2^+_1$ and $1/2^+_1$ states from the SM calculations
		for the Cl (magenta) and K (green) isotopes.
	}
\end{figure}

The odd-A $K$ isotopes show a similar behavior with respect to the 
lowest $3/2^+$ and $1/2^+$ 
states. The energy difference between the measured $3/2^+$ and $1/2^+$ states 
(one of which will be the ground state) 
is displayed in Figure~\ref{fig:oddA_Cl_K_isotopes} for isotones of Cl and K with 
neutron numbers from 20 to 32 (solid symbols connected by thin dashed line). 
In both cases we start with a $3/2^+$ ground state at $N=20$ with an inversion 
at higher neutron number, though for $K$ it happens at a larger neutron excess 
when the neutrons are 
filling the $1p_{3/2}$ orbital rather than $0f_{7/2}$ as for Cl isotopes. The 
energy differences as predicted by the shell model calculations are also shown by the 
solid lines. The calculations track the experimental observations well as far the 
inversion of the ground state from $3/2^+$ to $1/2^+$ goes for both, but the return back to 
a $3/2^+$ ground state is missed at the correct neutron number.
Considering that the spin assignments for the ground state for $^{41,43}$S are 
only tentative, a different picture could emerge if future results diverge from 
the current trends. It should be noted here that the occurrence of a $1/2^+$ 
ground state in the shell model calculations is a play between the effective single 
particle energies and deformation.

\begin{figure}
	\includegraphics[width=\columnwidth]{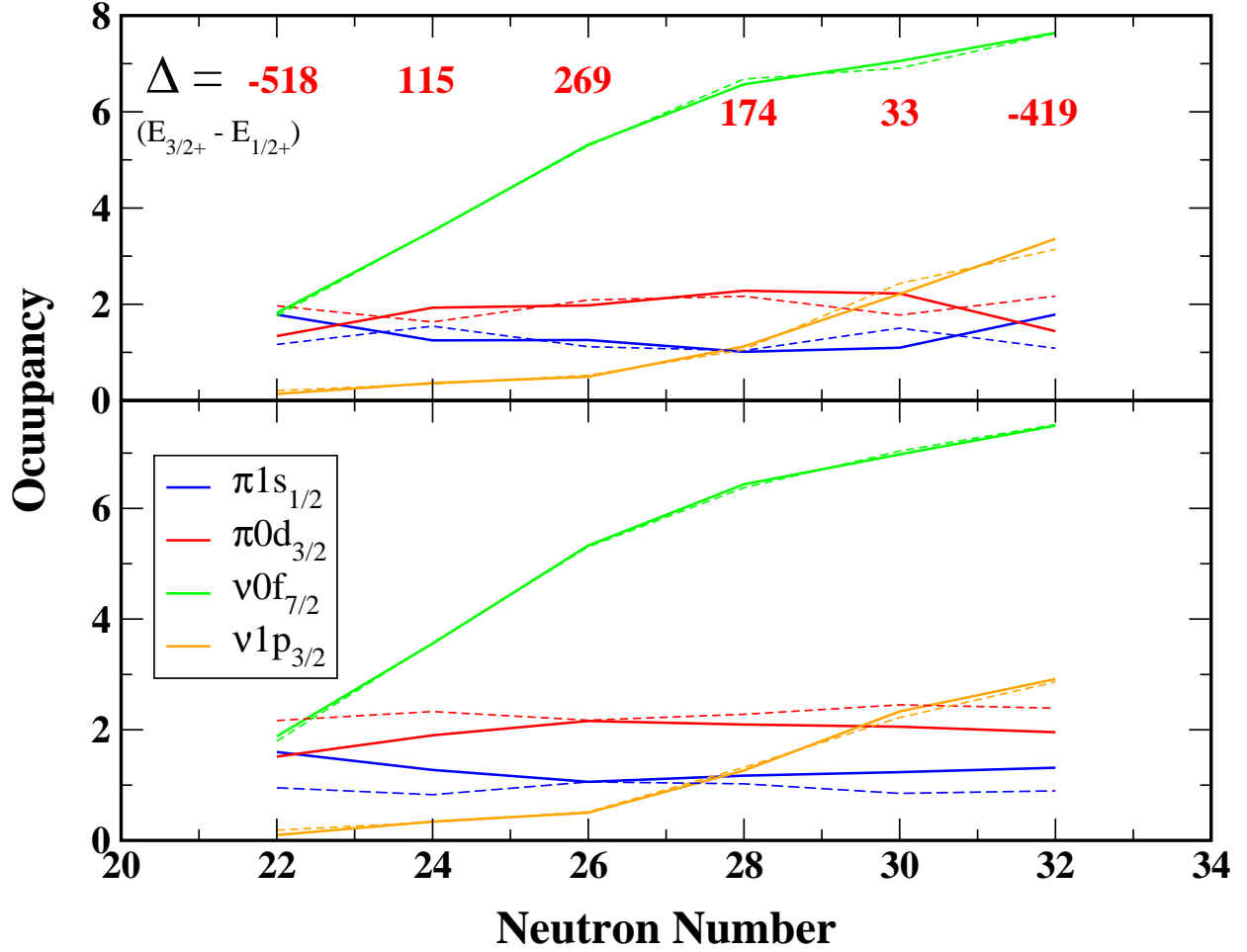}
	\caption{\label{fig:oddA_Cl_occupancy}
		The predicted occupancies of the proton $s_{1/2}$ and $d_{3/2}$ orbitals and neutron 
		$f_{7/2}$ and $p_{3/2}$ orbitals for the first four excited states in odd-A Cl 
		isotopes are displayed in the top panel. 
		These are orbitals where the valence nucleons would reside in these isotopes. 
		The top panel shows the occupancies for ground state (solid lines) and the first excited state
		(dashed line). For the ground state and the first excited state, the dominant occupancy of the 
		proton toggles between $0d_{3/2}$ and $1s_{1/2}$ orbitals for these isotopes which leads to the 
		determination of the ground state in the calculations. The energy difference calculated between 
		the $3/2^+_1$ and $1/2^+_1$ ($\Delta$) (in keV) from the shell model 
		calculations is also shown where the negative sign indicates a $3/2^+$ ground state. 
		The bottom panel displays the same for $5/2^+_1$ state (solid line) and $3/2^+_2$ 
		state (dashed line).
	}
\end{figure}

The occupancies of the first 4 excited states in Cl isotopes obtained from the 
shell model calculations are plotted in Figure~\ref{fig:oddA_Cl_occupancy}. 
The top panel shows the calculated occupancies for the ground state (solid line) 
and the first excited state (dashed line) as a function of neutron number.
$\Delta$ is the energy difference in keV between the calculated $3/2^+$ and $1/2^+$ states 
and one can see that the predictions are of a $1/2^+$ ground state (positive $\Delta$) 
for all isotopes with 
neutron number from 24 to 30 which is manifested in the 
occupancy of the $1s_{1/2}$ orbital equal to 1. This differs from the experimental observation 
where only $N=24$ and $26$ isotopes likely  have a $1/2^+$ ground state, though the calculated differences 
are small. Both the ground state and first excited state have similar occupancies for the 
$fp$ orbitals with the neutrons shared between $f_{7/2}$ and $p_{3/2}$ similarly.

The bottom panel of Figure~\ref{fig:oddA_Cl_occupancy} shows the occupancies of the excited 
$5/2^+_1$ (solid line) and $3/2^+_2$ (dashed line) states. The two states again show  similar 
occupation probabilities for the neutrons in the $fp$ orbitals with only differences in 
how the $\pi s_{1/2}$ and $\pi d_{3/2}$ orbitals are filled. 
The $3/2^+$ state consistently is formed by one proton in the 
$1s_{1/2}$ orbital and two in the $0d_{3/2}$ orbital. The $5/2^+$ state on the other hand has 
a more even distribution for the two orbitals, that is not such a pure configuration. 
The partial occupancies point towards a more collective nature for the $5/2^+_1$ state, 
which does follow the energetics of the $2^+$ states in the deformed $S$ core as was pointed 
out in Ref.~\cite{39_41Cl_negative}.

\subsection{\label{negative} $^{43,45}$Cl: Negative parity states}

The negative parity states in the even N ($ > 20$) isotopes of Cl are formed by the promotion 
of one particle from the $sd$ to the $fp$ shell. According to the shell model calculation based 
on the $SDPFSDG-MU$ interaction, the dominant configuration of the negative parity states 
is that of excitation 
of the unpaired proton from the $sd$ shell to the $0f_{7/2}$ or the $ 1p_{3/2}$ orbitals 
accompanied by some rearrangement of the $fp$ shell neutrons. 

\begin{figure*}
	\includegraphics[width=15cm]{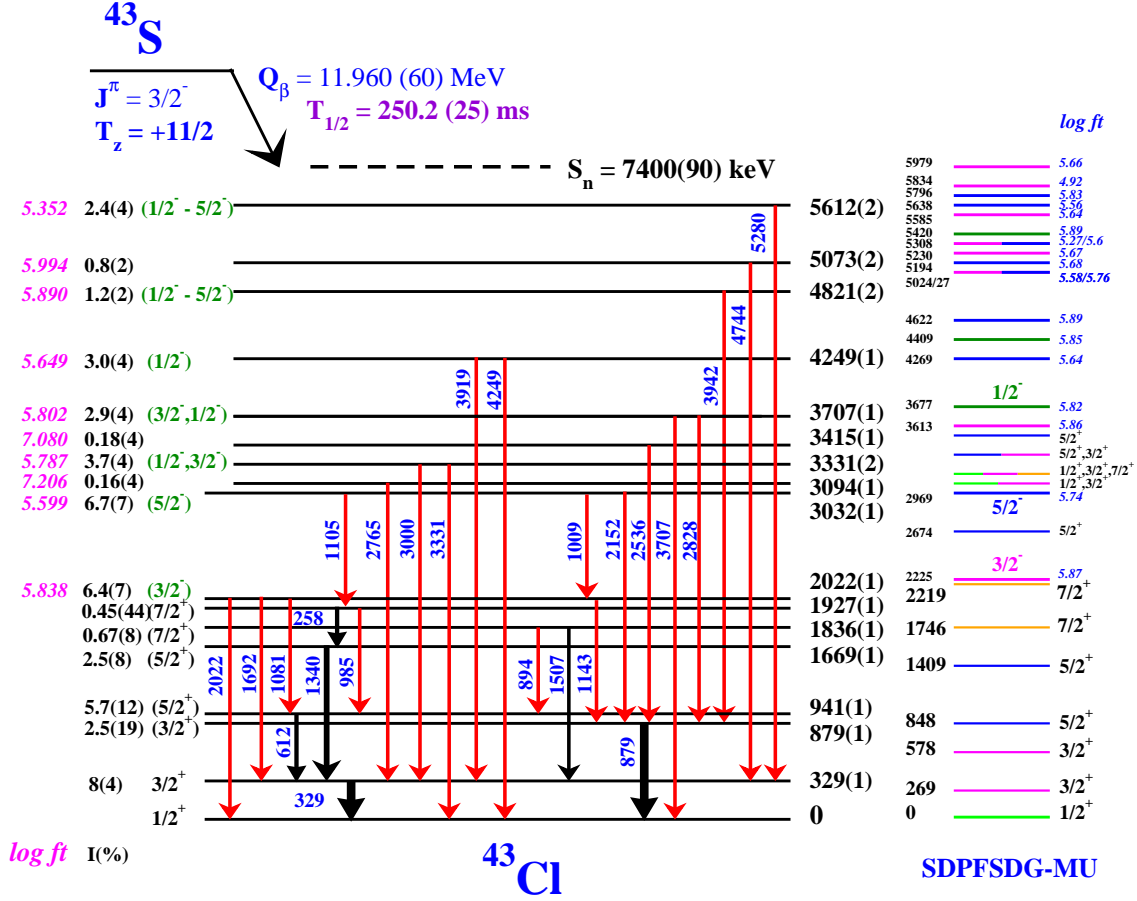}
	\caption{\label{fig:43S_levelscheme}
		Partial level scheme of $^{43}$Cl following the $\beta^-$-decay of $^{43}$S ( 
		$T_{1/2}$ = 250.2(25) ms and $Q_{\beta^-}$ = 11.960(60) MeV \cite{Q_beta}). The 
		transitions marked black were known before this work whereas red indicates 
		the new transitions added in this work. The absolute branching for each level 
		was calculated using the measured yields in the $\gamma$ transitions feeding in and out 
		of that state, measured $\gamma$ detection efficiency and  total number of implants 
		obtained from the fit to the decay curve. Using the $P_n$ value from Ref.~\cite{sorlin_Pn, marek} of 40(10)\% the feeding to the ground state can be estimated as 13(11)\% (not noted in the level scheme as $P_n$ 
  could not be measured for $^{43}$S in this work).   Log$ft$ values were calculated 
		using the log$ft$ calculator \cite{nndc_logft}. Alongside are the predictions of the 
		shell model calculation using the $SDPFSDG-MU$ interaction \cite{yoshida_website}. 
		The positive parity states represent $0p0h$ excitations while the negative parity 
		states are $1p1h$ excitations across the a major shell gap. For 
		protons it will be between $sd$ and $pf$ shells while neutrons can be excited 
		from the $sd$ to the $pf$ as well as from $pf$ to $sdg$ shells.
		Only selected $1/2^-$ (green), $3/2^-$ (magenta) and $5/2^-$ (blue) states 
		(with log$ft$ $<$ 6.00 and $E^* < 6$ MeV) along with the predicted log$ft$ 
  values 	are shown.
    }
\end{figure*}

\begin{figure*}
	\includegraphics[width=15cm]{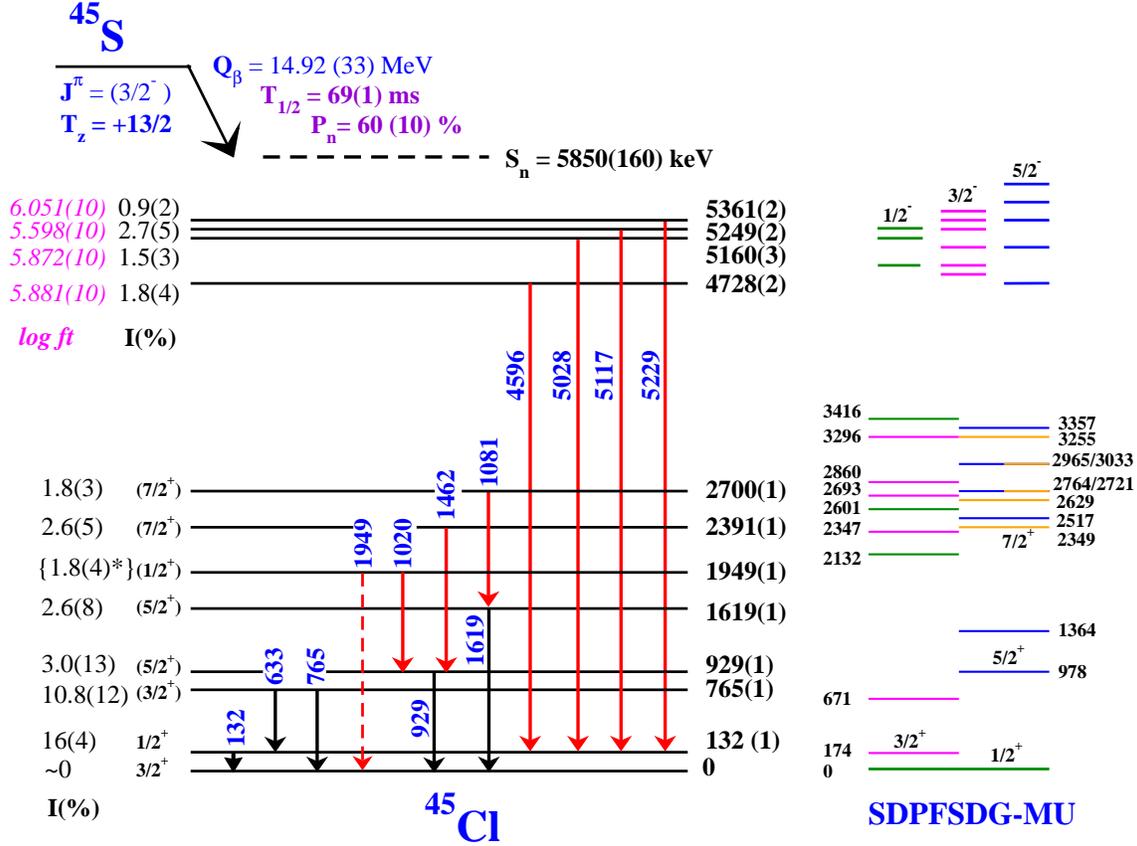}
	\caption{\label{fig:45S_levelscheme}
		Partial level scheme of $^{45}$Cl following $\beta$-decay of $^{45}$S 
		($T_{1/2}$ =  69(1) ms and $Q_{\beta^-}$ = 14.92(33) MeV \cite{Q_beta}). 
		The transitions marked black were known before whereas red indicates the 
		new transitions observed in this study. 
		The absolute branching for each level was calculated similar to that for 
  $^{43}$S decay.  
		As the intensity of the 1949 keV transition could not be obtained, the feeding 
  of the 1949-keV level is tentative(*).  Based on our measurements the ground state 
  feeding is consistent with zero.
		SM calculations similar to those for $^{43}$Cl  are shown alongside. Positive 
		parity states ($0p0h$) with spins of $1/2$, $3/2$, $5/2$ and $7/2$ below 3.5 MeV 
		are shown with different colors  identifying the spin values similar to Figure~\ref{fig:43S_levelscheme}.
		A few representative negative parity states ($1p1h$ excitations in the calculations) are 
		also shown. Only spins $1/2$, $3/2$ and $5/2$ with log$ft$ values below 6.00 
		in the energy range of the experimental states (above 3400 keV) are displayed for clarity. 
  The complete list can be	seen in Table~\ref{tab:45S_logft}.
	}
\end{figure*}

In $^{43}$Cl as shown in Figure~\ref{fig:43S_levelscheme} the 2022 keV state (log$ft$ = 5.88) 
is identified as the $3/2^-_1$ state consistent with the predicted state at 2225 keV with a 
log$ft$ of 5.87.  It has decays to the lower positive parity states via the 2022-, 1692-, 1143,
and 1081-keV transition with the ground state transition being the strongest 
(Table~\ref{tab:43Cl_gamma}). The 3032 keV level with decays to the $7/2^+_2$ and $3/2^-_1$ 
states is proposed  to be the $5/2^-_1$ state in agreement with the shell model calculations.
Both the 3331- and 3707-keV levels have comparable branches to the $1/2^+(gs)$ and $3/2^+_2$ 
and are candidates for the shell model states at 3613 keV ($3/2^-$) and 3677 keV ($1/2^-$) 
and are marked such in the level scheme. The 4249-keV level is proposed to be a $1/2^-$ state 
as a partner to the shell model state at 4409-keV with a comparable log$ft$ value. The levels 
above that are proposed to have negative parity but conjectures about spin values cannot be 
made from the current data.

\begin{figure}
	\includegraphics[width=\columnwidth]{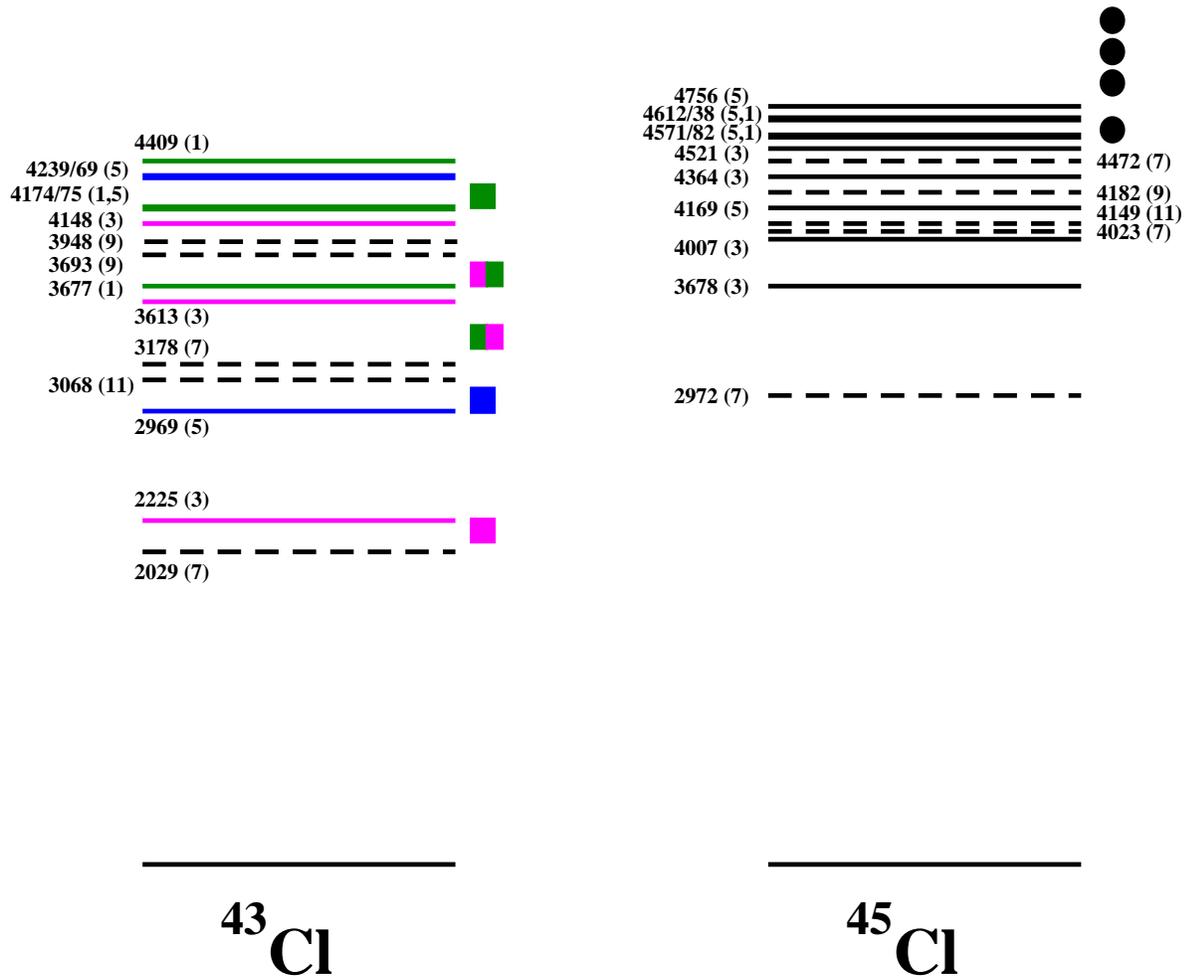}
	\caption{\label{fig:oddA_Cl_43_45}
		The first 15 negative-parity states calculated for $^{43}$Cl and $^{45}$Cl using
		the $SDPFSDG-MU$ interaction are shown. 
		The dashed lines are for spins $7/2^-$ and higher which 
		will not be populated in allowed GT decays while the solid lines show the lower spins.
		Excitation energy of the states in given in keV and the spin value as $2J$ (in parenthesis) 
		is displayed alongside. The solid symbols represent the experimental states with proposed 
		negative parity in the two cases.
		For $^{43}$Cl the $1/2^-, 3/2^-, 5/2^-$ are matched by colors consistent with 
		what is shown in Figure~\ref{fig:43S_levelscheme}. For $^{45}$Cl no spin values 
		have been assigned for the negative parity states and are all shown as black.
		See text for further discussion.
	}
\end{figure}

The first 15 negative parity states calculated for $^{43,45}$Cl are shown by solid and 
dashed lines in Figure~\ref{fig:oddA_Cl_43_45}. The solid lines are for spins 
$1/2^-, 3/2^-, 5/2^-$ which will be populated in the $\beta$-decay 
of the odd-A $S$ isotopes,$^{43,45}$S, while spins of $7/2^-$ and higher are shown as dashed. 
For both, the solid symbols represent the experimental states that are proposed to have 
negative parity. In 
the case of $^{43}$Cl we see an excellent agreement between the negative parity states observed here 
for the first time and the calculated ones. However, as discussed above the spin/parity 
assignment to the 
excited states shown in Figure~\ref{fig:43S_levelscheme} did use guidance from the shell model 
calculations along with the measured log$ft$ values and $\gamma$ branching. The experimental 
partners to the low spin states not only line up in excitation energy but the log$ft$ 
values are consistent too. This highlights the success of the shell model calculations in 
predicting the structure of both parent and the daughter equally well.

Compared to $^{43}$Cl, in $^{45}$Cl the negative parity states are predicted higher in energy, 
with $7/2^-$ as the first negative parity state at 2972 keV (equivalent state in $^{43}$Cl is 
at 2029 keV). The second state in both isotopes is a $3/2^-$; at 2225 keV in $^{43}$Cl while 
at 3678 keV in $^{45}$Cl.  For $^{43}$Cl the experimental state
at 2022 keV is in excellent agreement with the predictions with a log$ft$ value of 5.88, 
making the $3/2^-_1$ state one of the strongly populated states. Contrary to that, 
in $^{45}$Cl the predicted $3/2^-_1$ state at 3678 keV has a log$ft$ of 
6.61 (Table~\ref{tab:45S_logft}) and we do not have an experimental counterpart to that. 
Beyond the $3/2^-_1$ state the predicted negative parity states in $^{45}$Cl get very close in energy. 
The experimental states which have been tentatively assigned 
negative parities all lie above 4.7 MeV and there are counterparts in the shell model calculations 
(see again Table~\ref{tab:45S_logft}) with spins of 1/2, 3/2 or 5/2  though we cannot make a one 
to one correspondence for each spin. Compared to $^{43}$S decay the $B(GT)$ strength in $^{45}$S 
has moved to higher energies and is a reason for the large $\beta$-delayed neutron emission. 
A plausible explanation for this shift is that with a closed $N=28$ shell $^{45}$Cl may have a spherical 
configuration in contrast to the parent $^{45}$S which is likely to be deformed. States at 
higher excitation energy thus may provide a better overlap in that case and we may be seeing 
an influence of deformation in the $\beta$-decay strength distribution beyond the more prominent 
spin selection.

\subsection{\label{beta1n} $^{44}$Cl: Beta-delayed neutron emission}

\begin{figure}
	\includegraphics[width=\columnwidth]{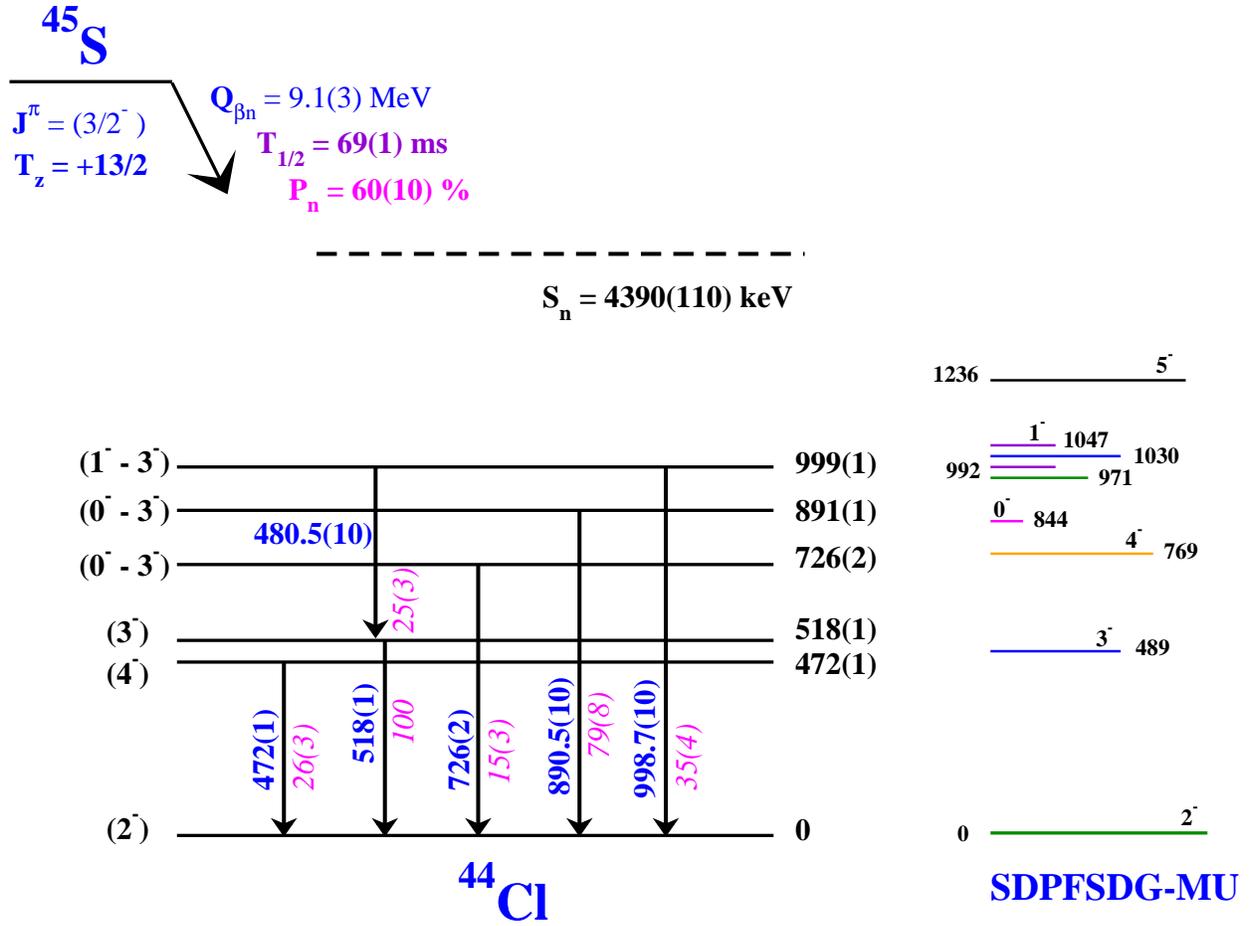}
	\caption{\label{fig:44Cl_levelscheme}
		Partial level scheme of $\beta$-delayed neutron daughter $^{44}$Cl. The number in 
		italics (magenta) alongside the energy of the transition (in keV) 
		represents the relative intensity with 
		the 518-keV transition taken as 100\%. Comparison with SM calculation (shown on the right) 
		and decay pattern have been used to suggest tentative spin-parities of the excited states.
	}
\end{figure}

Table~\ref{tab:45S_logft} lists the log$ft$ values calculated for the allowed $GT$ transitions 
in the $\beta$-decay of $^{45}$S (ground state $J^\pi$ of $3/2^-$) up to 7 MeV whereas the 
neutron separation energy in $^{45}$Cl is 5850(16) keV. 
We can see that above the neutron separation energy in $^{45}$Cl, there are several 
states with spins $1/2^-$, $3/2^-$ and $5/2^-$ which could be the intermediate states in the 
$\beta$-delayed neutron emission. Neutron emission from these would populate  
$0^-~\rm{to}~3^-$ states in the $\beta 1n$ daughter $^{44}$Cl if the neutron emitted has an
orbital angular momentum $\ell = 0$ which should be the most probable. 
The partial level scheme of the odd-odd $^{44}$Cl from the present study was shown in 
Figure~\ref{fig:44Cl_levelscheme} along with the low lying negative states predicted by the shell 
model calculations. An odd proton in the $0d_{3/2}$ orbital coupled to an odd neutron in the
$0f_{7/2}$ orbital for $^{44}$Cl 
will lead to a multiplet of states with spins from $2^-$ to $5^-$ whereas the other possibility of
proton in the $0d_{3/2}$ orbital and a neutron in the $1p_{3/2}$ orbital will lead to $0^-$ to $3^-$ 
spin-parity states. These states are very sensitive to the neutron-proton 
interaction in those orbitals and the low-lying states in $^{44}$Cl predicted by the SM 
calculations do correspond to these configurations (Figure~\ref{fig:44Cl_levelscheme}). 

The first excited state in $^{44}$Cl observed at 472 keV is likely a $4^-$ state based on the 
$E2$ transition strength deduced from the lifetime estimates in Ref.~\cite{riley_44Cl} for the 
same state. The shell model 
calculations produce the first $4^-$ state at 769 keV in reasonable agreement. The 518 keV 
state is likely a $3^-$ state in excellent agreement with the shell model state at 489 keV.
 If the 518 kev state is assigned a $3^-$ that will limit the spin-parity possibilities for the 999-keV excited state to $1^-, 2^-~or~3^-$ with decay branches to the 
 $3^-$ state and the $2^-$ ground state. The states at 726- and 891 keV are candidates for 
 states with $J^\pi$ = $0^-,1^-,2^-,3^-$ predicted by the shell model calculations  
 with similar  energies. 
 The generally good agreement between the states observed in the $\beta 1n$ decay 
 with the predicted low lying spectrum of $^{44}$Cl seems to corroborate the conjecture 
 of an $\ell = 0$ neutron being emitted from specific unbound 
 states in $^{45}$Cl. Neutron spectroscopy in the future will be greatly helpful 
 in understanding the $\beta$-delayed neutron emission process better.

\section{\label{conclude}Summary}

$\beta^-$ decay of the rare $^{43,45}$S isotopes around $N=28$ shell 
gap were studied at the NSCL using a 40x40 DSSD as the implant and $\beta$ detector, 
surrounded by a $\gamma$ array of 16 Clover detectors in close geometry. 
The half-lives were extracted from the fitting of decay curves for $^{43,45}$S
and are mostly consistent with the earlier work from GANIL. This was the 
first study of $\beta$-delayed $\gamma$ transitions for these isotopes 
and led to expanded level schemes for $^{43,45}$Cl with negative parity $1p1h$ 
states identified for the first time. We could provide one more argument in 
support for a $3/2^+$ ground state for $^{45}$Cl while $1/2^+$ for $^{43}$Cl.

Large scale shell model calculations using 
the $SDPFSDG-MU$ interaction provided a good reproduction of the experimental data. 
The shell model predicts the negative-parity states to be predominantly from 
proton excitation across the $Z=20$ shell gap for both $^{43,45}$Cl. 
The negative-parity states in $^{45}$Cl are found at least an MeV higher than what is seen
 for $^{43}$Cl. On the other hand, if the tentative state in $^{41}$Cl at 1475 keV is 
 identified with the SM predicted $7/2^-_1$ (1634 keV) then we also see an 
 upward trend in the occurrence of cross-shell excitations from $^{41}$Cl to 
 $^{43}$Cl.  Also the $\beta^-$ decay in $^{45}$S populates 
negative parity states higher up in the excitation spectrum. Both of these 
observations are likely related to moving towards the
 $N=28$ shell gap closure for $^{45}$Cl.

 Following the $\beta$-delayed neutron emission, low lying states in $^{44}$Cl 
 were also populated which are proposed to be 
of negative parity. The same states were seen earlier in-beam $\gamma$-ray 
spectroscopy and a few also in the direct $\beta$-decay of even-even $^{44}$S where they 
were populated via FF transitions. Tentative spin assignments have been made 
to these states for now. These states seem to be consistent with an $\ell =0$ 
neutron emission from unbound $1/2^-,3/2^-,5/2^-$ states in $^{45}$Cl.

$^{45}$S is the heaviest odd-A $S$ isotope, the $\beta^-$ of which has been studied 
until now. In our recent publication from this very data set we reported the 
$\beta^-$ decay of $^{46}$S which is the farthest from stability that can be currently 
reached for this element. 
The extension to $^{47}$S decay study should be possible with FRIB beams in the 
near future and will test the predictive powers of shell model calculations 
in this extremely neutron rich region close to the drip line.

\section{Acknowledgement}

We thank the NSCL operation team and the A1900 team specially Tom Ginter for 
the production and optimization of the secondary beam.  
This work was supported by the U.S. National Science Foundation under 
Grant Nos. PHY-2012522 (FSU), PHY-1848177 (CAREER);   
US Department of Energy, Office of Science, Office of Nuclear Physics
under award Nos. DE-SC0020451 (FRIB), DE-FG02-94ER40848 (UML), DE-AC52-07NA27344 (LLNL),
DE-AC02-06CH11357(ANL) and also by the US Department of Energy
(DOE) National Nuclear Security Administration Grant No. DOE-DE-NA0003906, 
the Nuclear Science and Security Consortium under Award No. DE-NA0003180.
YU and NS acknowledge KAKENHI grants (20K03981, 17K05433), “Priority Issue on
post-K computer” (hp190160, hp180179, hp170230) and
“Program for Promoting Researches on the Supercomputer Fugaku” (hp200130, hp210165). 
SY acknowledges JSPS KAKENHI Grant Number 22K14030 and KAKENHI grant 17K05433 
for supporting this work.

\bibliography{vandanatripathi_43_45S}


\begin{figure}
	\includegraphics[clip,width=12cm, height = 8cm]{Fig2a_43S_decaycurve.eps}
	\includegraphics[clip,width=12cm, height = 8cm]{Fig2b_43S_decaycurve_gamma.eps}
	\caption{\label{fig:43S_decaycurve}
		(a) Decay  curve for $^{43}$S from $\beta$-correlated implants
		within a grid of 9 pixels in the DSSD for 2 second correlation, 
		along with the fit used to extract 
		half-life and the initial activity.	Components of the fit are  
		(i) exponential decay of parent, $^{43}$S,	(ii)  exponential  growth  
		and  decay  of  daughter  nuclei,  $^{43}$Cl  ($\beta 0n$) and	$^{42}$Cl 
		($\beta 1n$),  and  (iii) exponential background. 
		Known half-lives were used  for the daughter nuclides  \cite{nndc}. 
		The half life  extracted is 256(5) ms in  agreement with the 
		evaluated value, 265(15) ms. 
		(b) Decay curve gated by the 329- and 879-keV transitions the two 
		lowest and strongest transitions in the daughter $^{43}$Cl 
		(see Figure~\ref{fig:43S_45S_gammas} and Figure~\ref{fig:43S_levelscheme}). 
		The half life extracted using 
		an exponential fit with a constant background is 250.2(25) ms.
	}
\end{figure}


\begin{figure}
	\includegraphics[clip,width=12cm, height = 8cm]{Fig5a_45S_decaycurve.eps}
	\includegraphics[clip,width=12cm, height = 8cm]{Fig5b_45S_decaycurve_gamma.eps}
	\caption{\label{fig:45S_decaycurve}
		(a) Decay  curve for $^{45}$S  from $\beta$-correlated implants 
		within a grid of 9 pixels in the DSSD, correlated for 1 second along 
		with the fit used to extract  half-life and  the initial activity. 
		Components of the fit are 
		(i) exponential decay of parent,  $^{45}$S, (ii)  exponential  growth  and  
		decay  of  daughter  nuclei,  $^{45}$Cl ($\beta 0n$) and $^{44}$Cl  
		($\beta$ 1n),  and  (iii) background.  
		Known half-lives were used  for the daughter nuclides  \cite{sorlin_Pn,soumik_45Cl,nndc}. 
		(b) Decay curve gated by the 132-keV transition which represents 
		the $1/2^+_1 \rightarrow 3/2^+ (gs) $ transition in daughter $^{45}$Cl.
		The extracted half-life values are 69(1) ms and 75(5) ms from the 
		two different conditions.
	}
\end{figure}


\begin{table}[b]
	\caption{\label{tab:43Cl_gamma}
		$\gamma$-ray energies, placements and efficiency corrected absolute intensities 
		(per 100 decay events) 
		of transitions in $^{43}$Cl from the  $\beta$ decay of $^{43}$S decay. }
	\begin{tabular}{ccc}
		\hline
		$E_\gamma$ (keV) ~~~&~~~ $E_{\it i}$ $\rightarrow$ $E_{\it f}$ (keV)~~~ & ~~~I (\%)\\
		\hline
		257.6(5)& 1927 $\rightarrow$ 1669 & 1.97(30)\\
		329.0(4)& 329 $\rightarrow$ 0 & 27.4(41)\\
		612.3(3)& 941 $\rightarrow$ 329 & 7.8(12)\\
		879.1(5)& 879 $\rightarrow$ 0 & 10.9(17)\\
		894.0(10)& 1836 $\rightarrow$ 941 & 0.39(6)\\
		985.0(10)& 1927 $\rightarrow$ 941 & 0.58(9)\\
		1009.4(5)& 3032 $\rightarrow$ 2022 & 0.91(14)\\
		1081.1(5)& 2022 $\rightarrow$ 941 & 1.15(18)\\
		1105.0(5)& 3032 $\rightarrow$ 1927 & 2.1(3)\\
		1143.0(3)& 2022 $\rightarrow$ 879 & 1.96(30)\\
		1340.2(3)& 1669 $\rightarrow$ 329 & 4.5(7)\\
		1506.5(6)& 1836 $\rightarrow$ 329 & 0.28(5)\\
		1692.0(5)& 2022 $\rightarrow$ 329 & 0.49(8)\\
		2022.0(8)& 2022 $\rightarrow$ 0 & 3.67(56)\\
		2152.1(5)& 3032 $\rightarrow$ 879 & 3.7(6)\\
		2536.0(10)& 3415 $\rightarrow$ 879 & 0.18(4)\\
		2765.3(10)& 3094 $\rightarrow$ 329 & 0.16(4)\\
		2828.0(6)& 3707 $\rightarrow$ 879 & 1.4(2)\\
		3000.0(5)& 3331 $\rightarrow$ 329 & 1.6(3)\\
		3331.0(5)& 3331 $\rightarrow$ 0 & 2.1(3)\\
		3707.0(10)& 3707 $\rightarrow$ 0 & 1.5(3)\\
		3919.4(10)& 4249 $\rightarrow$ 329 & 1.4(2)\\
		3942.0(10)& 4821 $\rightarrow$ 879 & 1.2(2)\\
		4249.0(20)& 4249 $\rightarrow$ 0 & 1.6(3)\\
		4744.0(20)& 5073 $\rightarrow$ 329 & 0.8(2)\\
		5280.0(20)& 5612 $\rightarrow$ 329 & 2.4(4)\\
		\hline		
	\end{tabular}
\end{table}


\begin{table}[b]
	\caption{\label{tab:45Cl_gamma}
		$\gamma$-ray energies, placements and efficiency corrected absolute intensities 
		(per 100 decay events) 
		of transitions in $^{45}$Cl from the  $\beta$ decay of $^{45}$S decay. }
	\begin{tabular}{ccc}
		\hline
		$E_\gamma$ (keV) ~~~&~~~ $E_{\it i}$ $\rightarrow$ $E_{\it f}$ (keV)~~~ & ~~~I (\%)\\
		\hline
		131.7(5)& 132 $\rightarrow$ 0 & 26(4)\\
		633.3(6)& 765 $\rightarrow$ 132 & 3.3(5)\\
		764.5(5)& 765 $\rightarrow$ 0 & 7.5(11)\\
		929.2(5)& 929 $\rightarrow$ 0 & 7.4(11)\\
		1020.1(10)& 1949 $\rightarrow$ 929 & 1.8(4)\\
		1081.2(10)& 2700 $\rightarrow$ 1619 & 1.8(3)\\
		1461.7(10)& 2391 $\rightarrow$ 929 & 2.6(5) \\
		1618.5(5)& 1619 $\rightarrow$ 0 & 4.4(7)\\
		1949.0(15)& 1949 $\rightarrow$ 0 & - \\
		4596.0(20)& 4728 $\rightarrow$ 132 & 1.8(4)\\
		5028.4(30)& 5160 $\rightarrow$ 132 & 1.5(3)\\
		5117.1(20)& 5249 $\rightarrow$ 132 & 2.7(5)\\
		5229.0(20)& 5361 $\rightarrow$ 132 & 0.9(2)\\
		\hline		
	\end{tabular}
\end{table}


\begin{table*}[b]
	\caption{\label{tab:Cl_iso_branching}
		Comparison of the calculated decay branches of the first and second $5/2^+$ states to the 
		ground state and the first excited state in $^{39-49}$Cl isotopes. As discussed in the text for 
		$^{41,43}$Cl the ground state is $1/2^+$ with a $3/2^+$ first excited state and the opposite 
		for other isotopes. The shell model predictions for $B(L)$ values and the experimental 
		excitation (SM calculated excitation) energies were used to calculate the decay rates 
		for the two $5/2^+$ states. In the calculations, the EM transition matrix elements are 
		evaluated with the effective $g$-factors: $g_\ell ^\pi$ = 1.15, $g_\ell ^\nu$ = -0.15  
		and $g_s = 0.85 * g_s^{bare}$.
		The effective charges used are 1.35e ($\pi$) and 0.35e ($\nu$). 
		The numbers are quoted for a 100\% population  of the said state.
	}
	\begin{tabular}{|c|cc|cc|}
		\hline
		~{Cl isotope}~ &~~ {$5/2^+_1$ $\rightarrow$ $3/2^+_1$}~~ & ~~{$5/2^+_1$ $\rightarrow$ $1/2^+_1$} ~~& ~~{$5/2^+_2$ $\rightarrow$ $3/2^+_1$}~~ & ~~{$5/2^+_2$ $\rightarrow$ $1/2^+_1$}\\
		\hline\hline
		$^{39}$Cl & 99.7 (100) & 0.3 (0) & 99.2 (100) & 0.8 (0)\\	   
		\hline	
		$^{41}$Cl & 97.7 (98.4) & 2.3 (1.6) & - (100) & - (0) \\	   
		\hline
		$^{43}$Cl & 89.6 (92.4) & 10.4 (7.6) & 98.2 (98.7)  & 1.8 (1.3) \\	   
		\hline	
		$^{45}$Cl & 97.2 (88.7) & 2.8 (11.3) & 98.1 (96.9) & 1.9 (3.1) \\	   
		\hline   
		$^{47}$Cl & - (64.4) & - (35.6) & - (98) & - (2) \\	   
		\hline    
		$^{49}$Cl & 95.3 (93.8) & 4.7 (6.2) & 56.5 (59) & 43.5 (43) \\	   
		\hline  
	\end{tabular}
\end{table*}


\begin{table}[b]
	\caption{\label{tab:45S_logft}
		Shell model predictions for the  $\beta ^-$ decay of $^{45}$S with a ground state $J^\pi$ = $3/2^-$.
		Excited states up to 7~MeV in $^{45}$Cl with spins of $1/2^-$, $3/2^-$, and $5/2^-$ expected to be 
		populated in allowed $GT$ decay are given with the calculated log$ft$ values. }
	\begin{tabular}{|cc|cc|cc|}
		\hline
		\multicolumn{2}{|c|}{$3/2^-$ $\rightarrow$ $1/2^-$} & \multicolumn{2}{c|}{$3/2^-$ $\rightarrow$ $3/2^-$} & \multicolumn{2}{c|}{$3/2^-$ $\rightarrow$ $5/2^-$}\\
		\hline \hline
		{$E_{level}$ (keV)}~ &~ {log$ft$}& {$E_{level}$ (keV)}~ & ~{log$ft$}&{$E_{level}$ (keV)}~ & ~{log$ft$}\\		
		\hline
		4583 & 8.26 & 3678 & 6.61 & 4.170 & 6.97 \\	
		4369 & 7.19 & 4008 & 6.83 & 4572 & 6.18 \\
		4956 & 5.67 & 4365 & 6.73 & 4612 & 7.76 \\
		5209 & 6.88 & 4522 & 6.17 & 4757 & 5.93 \\
		5253 & 5.70 & 4873 & 5.48 & 4959 & 7.07 \\
		5396 & 6.13 & 4968 & 5.53 & 5022 & 6.28 \\
		5567 & 9.20 & 5161 & 5.48 & 5187 & 5.40 \\
		5746 & 7.90 & 5301 & 5.72 & 5377 & 5.55 \\
		6003 & 6.43 & 5396 & 5.33 & 5515 & 5.76 \\
		6089 & 6.44 & 5501 & 5.31 & 5712 & 5.38 \\
		6355 & 6.38 & 5670 & 5.75 & 5900 & 4.97 \\
		6483 & 6.22 & 5928 & 5.43 & 5991 & 5.37 \\
		6699 & 6.03 & 6074 & 5.65 & 6281 & 5.53 \\
		6997 & 5.98 & 6382 & 5.90 & 6532 & 5.37 \\
		& &  6723 & 6.14 & 6797 & 5.30 \\
		& & 6992 & 5.66 & 7029 & 5.35 \\
		
		\hline		
	\end{tabular}
\end{table}

\end{document}